\documentclass[%
 reprint,
showpacs,preprintnumbers,
 amsmath,amssymb,
 aps,
]{revtex4-1}

\usepackage{graphicx}
\usepackage{caption,subcaption}
\usepackage{dcolumn}
\usepackage{bm}

\begin{document}

\preprint{APS/123-QED}

\title{Helicity in axisymmetric vortex breakdown}

\author{Manjul Sharma}
\author{Sameen A}%
 \email{sameen@ae.iitm.ac.in}
\affiliation{%
 Department of Aerospace Engineering, IIT Madras, Chennai 600036, India
}%

\date{\today}

\begin{abstract}
Vortex breakdown phenomena in the axial vortices is an important feature which occurs frequently in geophysical flows (tornadoes and hurricanes)
and in engineering flows (flow past delta wings, Von-Kerman vortex dynamo).
We analyze helicity for axisymmetric vortex breakdown and propose a simplified formulation.
For such cases, negative helicity is shown to conform to the vortex breakdown.
A model problem has been analyzed to verify the results.
The topology of the vortex breakdown is governed entirely by helicity density in the vertical plane.
Our proposed methodology may be regarded as the prototype for identifying and characterize the breakdowns/eye in more complicated large-scale
flows such as tornadoes/hurricanes.
\end{abstract}

\maketitle
Axial vortices are ubiquitous in engineering and nature.
Some examples are flow in a rotating pipe, flow past delta wings, swirl type combustion chambers, bio-reactors, day-to-day
small scale flows like bathroom sink, stirring of coffee in a cup, and large scale geophysical flows such as tornadoes.
Axial vortices are prone to a phenomena called `vortex breakdown', first observed by Peckham \& Atkinson \cite{peckham1957} in flow past delta wings.
It is characterized by appearance of one or more stagnation points
on or near the axis of the vortex followed by either a recirculatory bubble or a spiral~\cite{leibovich1978,escudier1984,escudier1988,delery1994}.
Observations as early as in 1787 by Michaud~\cite{michaud1787} has documented schematics showing breakdowns in tornadoes and waterspouts.
These breakdowns occur at different stages of a tornado~\cite{ipauley1988}.
In hurricanes the vortex breakdown is referred as `eye', which occurs in their mature stage.
While formation of eye is fundamental to hurricanes, little is known about the formation process of the eye.
Above examples emphasize the underlying importance of identifying and understanding the phenomena of vortex breakdown.

To study the vortex breakdown phenomena we have used a model problem that generates the bubble-type vortex breakdown.
The model is the flow inside a circular cylinder with top rotating lid.
The vortex breakdown is generated inside a circular cylinder with top rotating lid.
This flow exhibits a bubble type vortex breakdown, which is characterized by the appearance of stagnation
points followed by a recirculatory region along the axis.
Two non-dimensional parameters that govern the flow are: (i) aspect ratio, $\Gamma=H/R$, where, $H$ and $R$ are the height and radius of the cylinder
and (ii) Reynolds number, $Re=\Omega R^{2}/\nu$, where, $\Omega$ is the rate at which the lid is rotated and $\nu$ is the kinematic viscosity.
By varying both the parameters it is possible to obtain one, two, and even three breakdown bubbles.
Escudier~\cite{escudier1984} has provided a map of the number of breakdown bubbles in $\Gamma-Re$ plane.
There have been many attempts to understand the underlying phenomena and the onset conditions behind the vortex breakdown bubble.
Benjamin~\cite{benjamin1962,benjamin1967} has explained vortex breakdown as a hydraulic jump from a supercritical state to a subcritical state.
Escudier~\cite{escudier1983} have considered vortex breakdown as a two stage transition.
Wang \& Rusak~\cite{wang1997} and Shtern \& Hussain~\cite{shtern1999} have explained vortex breakdown as a fold catastrophe that
occurs in rotating pipe flow.
Brown \& Lopez\cite{brown1990} have discussed the conditions under which the vortex breakdown occurs and proposed that
generation of negative azimuthal vorticity is an essential feature of the vortex breakdown.
This paper does not find any exception to this condition, however, the generation of the negative azimuthal vorticity is
found not to be exclusive to the vortex breakdown as it is generated in the flow even when the breakdown does not occur.

In this study we are using helicity density to investigate the vortex breakdown.
Helicity is regarded as one of the three important variables in three-dimensional flows, the other two being energy and the enstrophy.
It is also has been used extensively to analyze the stability and the energy propagation in atmospheric flows~\cite{elting1985}.
Lilly~\cite{lilly1986} has shown that supercell thunderstorms can be categorized with high helicity.

We have obtained a formulation of helicity simplified for axisymmetric flows and have applied to vortex breakdown generated by a model problem.
Helicity density is directly related to the local topology of the flow~\cite{moffatt1992} and represents the orientation of the
velocity vector $\mathbf{V}$ and the vorticity vector $\bm{\omega}$.
Possible relation of helicity with the vortex breakdown was first mentioned briefly by Moffatt \& Tsinober \cite{moffatt1992}.
They argued that vortex breakdown is a result of change in the topology of the flow and hence,
helicity density may be an appropriate parameter to characterize the breakdown.
The fact that the helicity density changes sign across a separation or reattachment line, makes it a suitable parameter
to predict the location of the vortex breakdown by locating the stagnation points on the axis.

Three-dimensional incompressible Navier-Stokes equations are solved numerically in cylindrical coordinate system $\left(\theta, r, z \right)$.
    \begin{equation}\label{mom}
     \frac{\partial \mathbf{V}}{\partial t} + \left(\mathbf{V}\cdot \nabla \right)\mathbf{V}=-\nabla p + \frac{1}{Re}\nabla^{2}\mathbf{V}
    \end{equation}
Here, $\mathbf{V}=\left\langle u_{\theta}, u_{r}, u_{z} \right\rangle$ is velocity and $p$ is pressure.
The non-dimensionalization is done using $R$ as the length scale and $\Omega R$ as the velocity scale.
A finite-difference method employing fractional-step algorithm \cite{verzicco1996} is used to solve the above equations.
The boundary conditions at lower wall and side wall are $ u_{\theta}=u_{r}=u_{z}=0$, and at upper wall are $u_{\theta}={r}/{R}, u_{r}=u_{z}=0$.
Fully staggered arrangement of variables has been used to achieve coupling between the pressure and the velocity.
A grid independent study has been carried out to find the most suitable grid for the cases that are considered in this study.
The solution for $Re=3400$ is simulated for three different grids which is the highest $Re$ reported in this paper.
The results are checked for consistency with both constant CFL number and constant time-step $\left(\Delta t\right)$ calculations.
Based on these results, a grid of size $\left(N_{\theta} \times N_{r} \times  N_{z} : 257 \times 129 \times 257 \right)$ is used for $\Gamma=2.5$ in this study.
The time step for the integration is $\Delta t=10^{-3}$.

\begin{figure}
\centering
  \begin{subfigure}[b]{0.32\columnwidth}
    \centering
        \includegraphics[width=\textwidth]{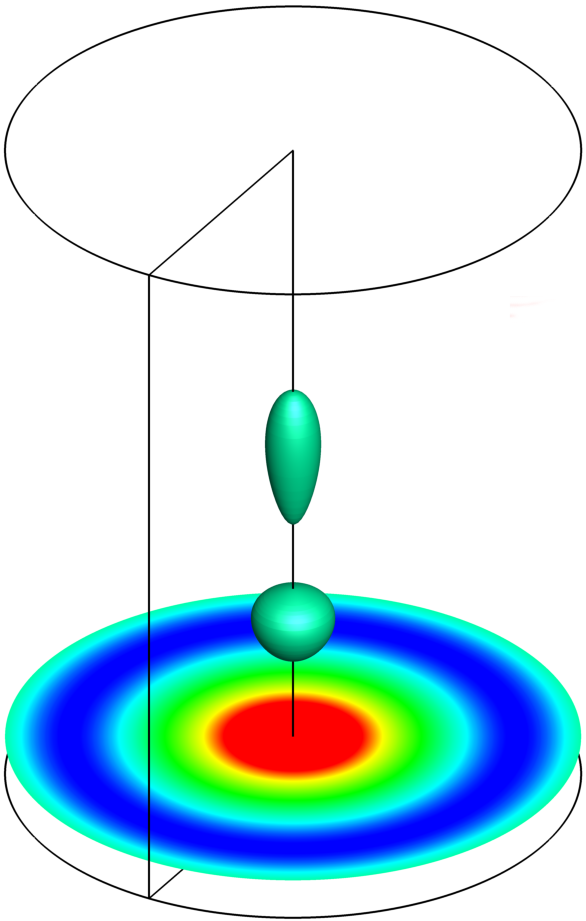}
        \caption{}
        \label{w_iso_re2200}
        \end{subfigure}\hfill
        \begin{subfigure}[b]{0.32\columnwidth}
        \centering
        \includegraphics[width=\textwidth]{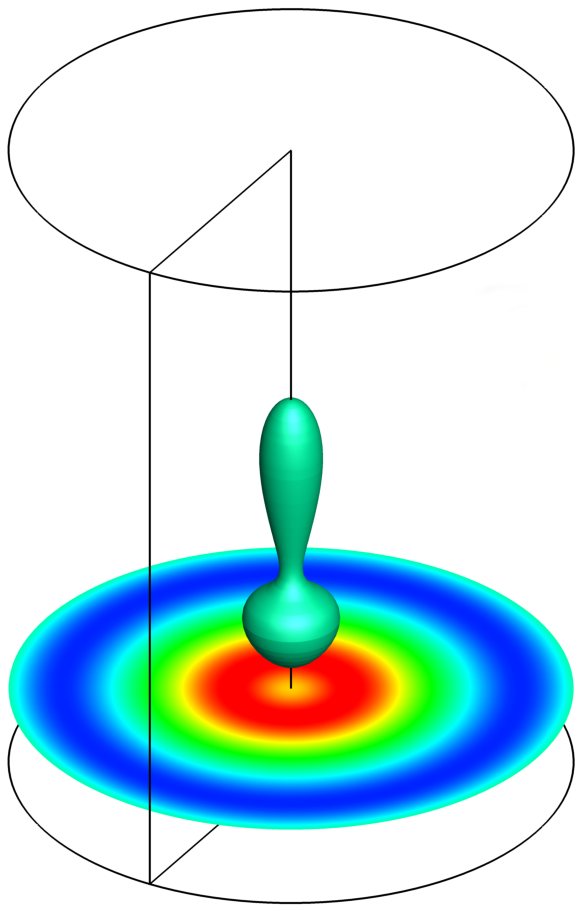}
        \caption{}
        \label{w_iso_re2494}
        \end{subfigure}\hfill
        \begin{subfigure}[b]{0.32\columnwidth}
        \centering
        \includegraphics[width=\textwidth]{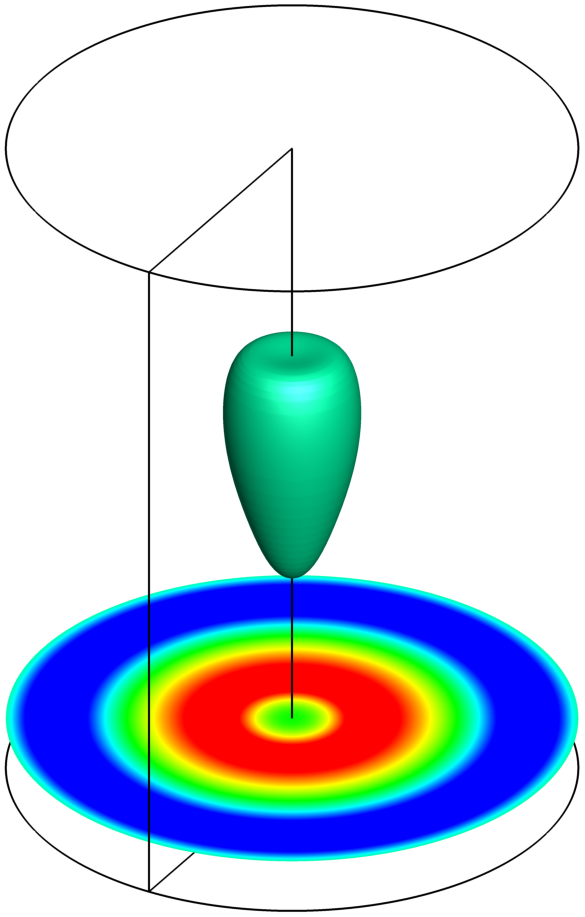}
        \caption{}
        \label{w_iso_re2700}
        \end{subfigure}\hfill
        \begin{subfigure}[b]{0.8\columnwidth}
        \centering
        \includegraphics[width=\textwidth]{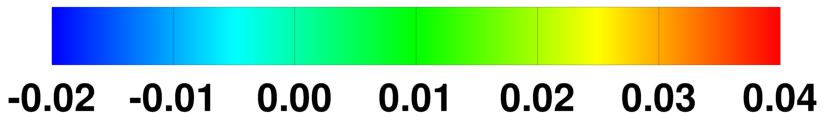}
        \label{}
        \end{subfigure}%
\caption{Contours of $u_{z}$ in $z-$plane at $z=0.1$ and iso-surface of $u_{z}=0$ for (a) $Re=2200$, (b) $Re=2494$, and (c) $Re=2700$ at an instant for $\Gamma=2.5$}
\label{flow2.5}
\end{figure}

The flow has been simulated for various aspect ratios but mainly the results for $\Gamma=2.5$ have been discussed and described in detail.
This aspect ratio has been examined well experimentally (\cite{escudier1984,fujimura2001}) as well as computationally (\cite{lopez1992,stevens1999,blackburn2000}).
Results for other aspect ratios are presented and discussed as and when required.
All the results presented in this study are in the axisymmetric regime of the flow.
For aspect ratio $\Gamma=2.5$, at $Re=2200$, two steady axisymmetric breakdown bubbles are obtained \cite{escudier1984}, see Fig. \ref{w_iso_re2200}.
There are 4 distinct stagnation points at the axis.
The iso-surface of $u_{z}=0$ is used to represent the structure and shape of the breakdown bubble as was used by Serre \& Bontoux \cite{serre2002}.
As the Reynolds number is increased to $2494$ both the bubbles merge leaving only two stagnation points on the axis as shown in the Fig. \ref{w_iso_re2494}.
At $Re=2700$ the flow becomes unsteady but remains axisymmetric and the breakdown bubble starts moving along the
axis periodically this observation was also reported by \cite{lopez1992}.
Fig. \ref{flow2.5} shows these transitions in the flow for $\Gamma=2.5$ as $Re$ is increased.
Brown and Lopez \cite{brown1990} have shown that for axisymmetric vortex breakdowns, the flow away from the breakdown bubble and the side wall can be assumed inviscid.
Under this assumption they have shown that in order to have the stagnation points on the axis and hence the breakdown bubble,
generation of the negative azimuthal vorticity is required in the vicinity of the axis.
\begin{figure}
\centering
  \begin{subfigure}[b]{0.46\columnwidth}
  \centering
  \includegraphics[width=\textwidth]{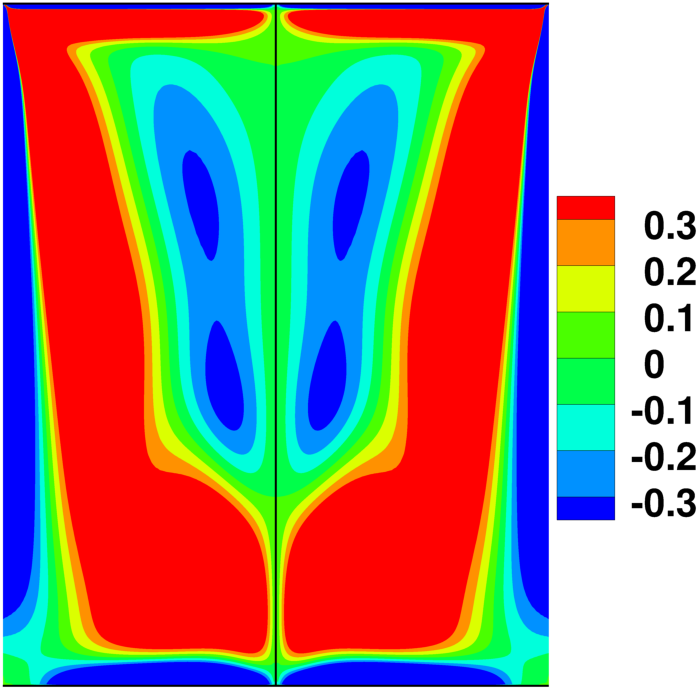}
  \caption{}
  \label{tvort1600}
  \end{subfigure}\hfill
  \begin{subfigure}[b]{0.49\columnwidth}
  \centering
  \includegraphics[width=\textwidth]{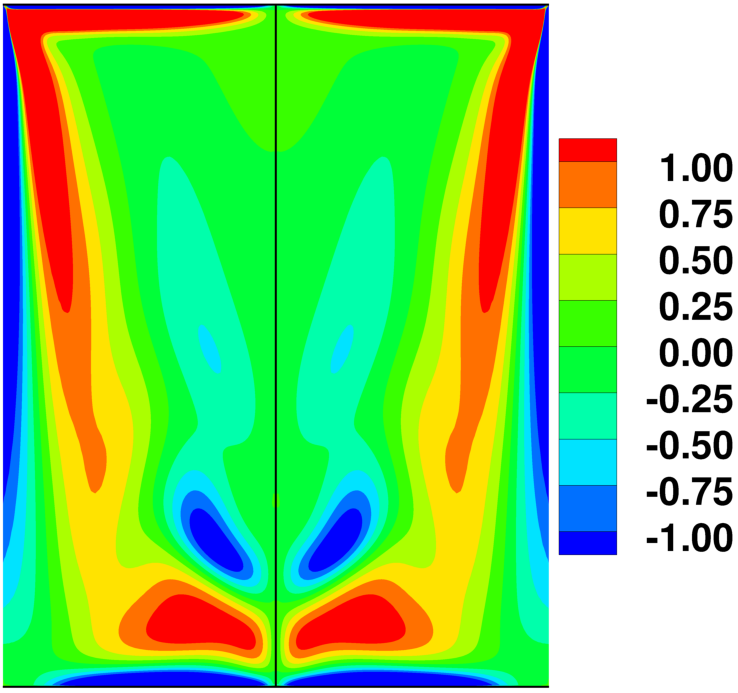}
  \caption{}
  \label{tvort2200}
  \end{subfigure}
\caption{Contours of azimuthal vorticity in $rz$ plane for $\Gamma=2.5$, (a) $Re=1600$ and (b) $Re=2200$.}
\label{vort2.5}
\end{figure}
\begin{figure}
\centering
      \includegraphics[width=0.8\columnwidth]{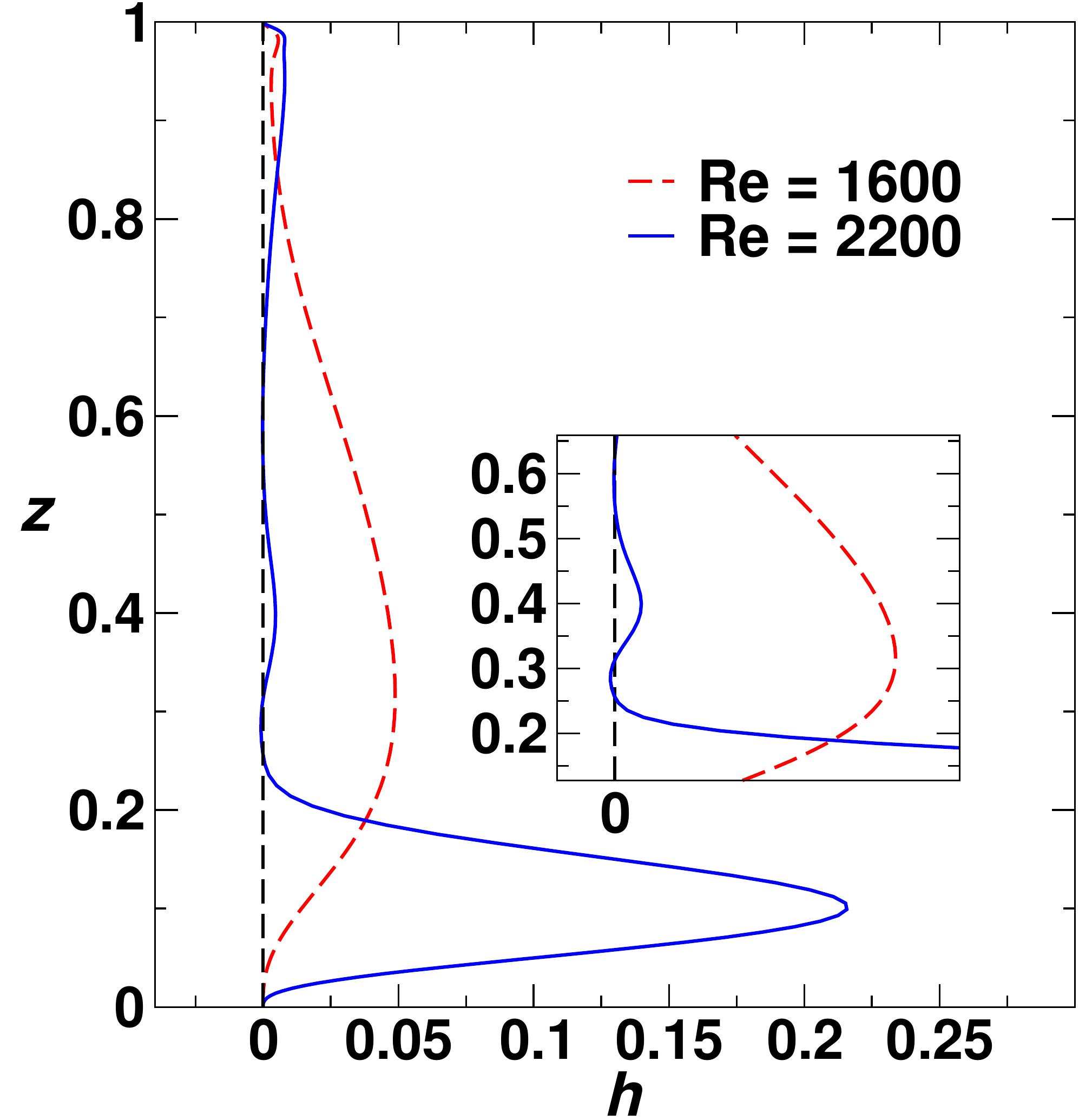}
\caption{Variation of helicity density, $h$, along the axis $\left(r=0 \right)$ for Figs. \ref{tvort1600} and \ref{tvort2200}.}
      \label{helicity2.5}
\end{figure}
The Fig. \ref{vort2.5} shows the contours of azimuthal vorticity for $\Gamma=2.5$.
For $Re=1600$ the flow does not have any breakdown bubble and Fig. \ref{tvort1600} shows negative azimuthal vorticity in the vicinity of the axis.
For $Re=2200$, which has two distinct vortex breakdown bubbles , Fig. \ref{tvort2200} also shows negative azimuthal vorticity
in the vicinity of the axis.
This indicates that even though negative azimuthal vorticity generation is essential to vortex breakdown,
its mere presence does not guarantee the vortex breakdown.

We analyze helicity density of the flow with axisymmetric vortex breakdown.
The helicity density is defined as, $h=\textbf{V}\cdot\bm{\omega}$.
Fig. \ref{helicity2.5} shows variation of $h$ along the axis of the cylinder for $Re=1600, 2200$ for $\Gamma=2.5$.
In case of $Re=1600$, since there is no stagnation point on the axis, the helicity density does not change sign along the axis.
While in the case of $Re=2200$, there are four stagnation points on the axis and Fig. \ref{helicity2.5} shows that helicity
changes sign exactly at these four points.
It can be seen in the Fig. \ref{helicity2.5} that between the stagnation points, where the breakdown bubbles are present, helicity is negative.
It has been observed that the helicity is negative inside and in the near vicinity of the breakdown bubbles and is enveloped by positive helicity.
This behaviour of the helicity can be easily understood at least at the axis of the cylinder.
The breakdown bubble is a surface where the circulation of the flow changes the direction as one moves radially in to the bubble or out of the bubble.
Moreover, at the axis, $u_{r}=0$ for the axisymmetric case.
The implications of these facts are that there is only axial flow present at the axis.
Apart from this region, the negative helicity occurs adjacent to the rotating lid.
The bulk clockwise circulation results in an upward flow outside the breakdown bubble and
a counterclockwise circulation within the breakdown bubble results in a downward flow within the breakdown bubble at the axis.
The evolution equation of helicity density, $h$, is,
\begin{eqnarray}\label{hel_transport}
\frac{\partial h}{\partial t}+u_{j}\frac{\partial h}{\partial x_{j}}=&-&\omega_{i}\frac{\partial p}{\partial x_{i}}+\omega_{j}\frac{\partial}{\partial x_{j}}\left(\frac{u^{2}_{i}}{2} \right)\\ \nonumber
&+&\frac{1}{Re}\left(\frac{\partial ^{2}h}{\partial x^{2}_{j}}-2\frac{\partial u_{i}}{\partial x_{j}}\frac{\partial \omega_{i}}{\partial x_{j}} \right)
\end{eqnarray}
We integrate the equation \ref{hel_transport} in the whole domain.
The second term, after integration, which is simplified near the bottom and the top plate is,
\begin{equation}\label{hel_injection}
\int_{S} \hat{n}\cdot\left(-h \mathbf{V} \right)dS\approx\int_{S}u_{z}hdS
\end{equation}
The term that shows that the helicity that is generated at the top rotating plate is injected in the bulk flow by the axial component of the velocity.
In other words, the equation \ref{hel_injection} shows that the helicity is transported by $u_{z}$ in the flow from the top plate.
This helicity gets concentrated in the vicinity of the axis where the breakdown occurs.

\begin{figure}
    \begin{subfigure}[b]{0.32\columnwidth}
      \centering
      \includegraphics[width=\textwidth]{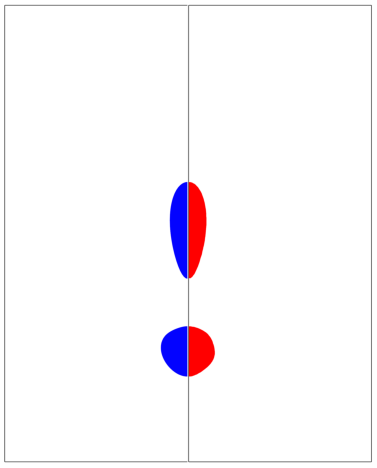}
     \caption{}
      \label{2200_t1}
    \end{subfigure}\hfill
    \begin{subfigure}[b]{0.325\columnwidth}
      \centering
      \includegraphics[width=\textwidth]{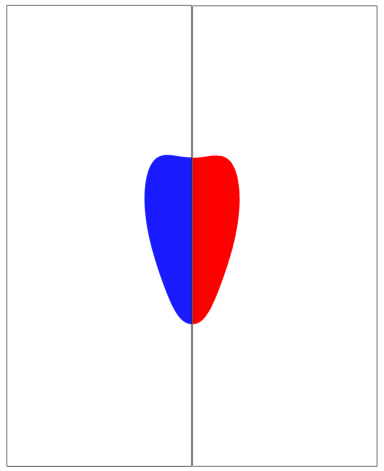}
      \caption{}
      \label{3000_t1}
    \end{subfigure}\hfill
    \begin{subfigure}[b]{0.325\columnwidth}
      \centering
      \includegraphics[width=\textwidth]{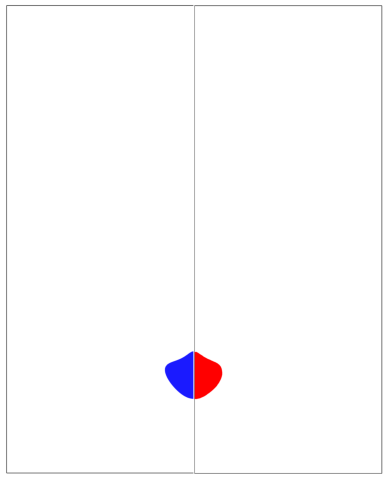}
      \caption{}
      \label{3000_t3}
    \end{subfigure}
\caption{Comparison of helicity in $rz$ plane with the breakdown bubbles for (a) $Re=2200$, $\Gamma=2.5$, (b) and (c) $Re=3000$, $\Gamma=2.5$ at two different instants. In each figure, the left half (blue) shows the breakdown bubble defined by the contours of $u_{z}=0$ and right half (red) shows the negative contours of helicity in $rz$ plane.}
\label{helicity_comparision}
\end{figure}

The role of helicity in relation with the vortex breakdown can be understood with the help of the axisymmetric flows.
For axisymmetric flows the formulation of the helicity can be simplified as follows.
The flow can be assumed as two-dimensional three component (2D3C) flow where the $\theta$ component of the velocity can be thought of as a passive component in a two-dimensional $rz$-plane flow~\cite{biferale2017}.
Full three-dimensional flow in such a case can be represented as,
\begin{equation}
\mathbf{u}=\mathbf{V}^{2D}+\bm{\phi}
\end{equation}
Here, $\mathbf{V}^{2D}$ is the two-dimensional part and $\bm{\phi}$ is the passive part given by,
\begin{equation}
\mathbf{V}^{2D}=\left\langle 0, u_{r}, u_{z} \right\rangle; \bm{\phi}=\left\langle u_{\theta}, 0, 0 \right\rangle
\end{equation}
Evolution of these two velocity components is governed by the following equations,
\begin{equation}
\frac{\partial \mathbf{V}^{2D}}{\partial t}+\left(\mathbf{V}^{2D} \cdot \nabla \right)\mathbf{V}^{2D} = -\nabla p + \nu \nabla ^{2} \mathbf{V}^{2D}
\end{equation}
\begin{equation}
\frac{\partial \bm{\phi}}{\partial t}+\left(\mathbf{V}^{2D} \cdot \nabla \right) \bm{\phi} = \nu \nabla ^{2} \bm{\phi}
\end{equation}
Vorticity fields associated with these velocities are $\bm{\omega}$ and $\bm{\omega}^{\phi}$, which are found as follows,
\begin{equation}
\bm{\omega}=\left\langle \frac{\partial u_{r}}{\partial z} - \frac{\partial u_{z}}{\partial r}, 0, 0 \right\rangle; \bm{\omega}^{\phi}=\left\langle 0, -\frac{\partial u_{\theta}}{\partial z}, \frac{u_{\theta}}{r}+\frac{\partial u_{\theta}}{\partial r} \right\rangle
\end{equation}
The helicity, $h_{r,z}$ in the $rz$ plane, which is associated with $\mathbf{V}^{2D}$ is then given by,
\begin{equation}\label{hrz_eqn}
h_{r,z}=\mathbf{V}^{2D} \cdot \omega^{\phi}=\underbrace{-u_{r} \frac{\partial u_{\theta}}{\partial z}}_{I} + \underbrace{\frac{1}{r} u_{z} u_{\theta} + u_{z} \frac{\partial u_{\theta}}{\partial r}}_{II}
\end{equation}
Similarly, out of the plane helicity, $h_{\theta}$, is given by,
\begin{equation}
h_{\theta}={u}_{\theta}\mathbf{e^{\theta}} \cdot \omega=u_{\theta}\left( \frac{\partial u_{r}}{\partial z} - \frac{\partial u_{z}}{\partial r} \right)
\end{equation}
For an axisymmetric case, it is the helicity in the $rz$ plane that determines the flow topology completely as it represents the alignment of $\mathbf{V}^{2D}$ vector
with respect to $\bm{\omega}^{\phi}$ vector in $rz$ plane.
Fig. \ref{2200_t1} shows contours of $h_{r,z}$ for $\Gamma=2.5$, $Re=2200$ respectively
compared against the breakdown bubbles which are represented by the surface of $u_{z}=0$ in the $rz$ plane.
For each right half of the Fig. \ref{helicity_comparision}, the values below $0$ are shown in color and white region shows the values above 0.
It can be seen that the contours of negative values of $h_{r,z}$ exactly match with the breakdown bubbles.
In fact, the region of negative $h_{r,z}$ along the axis of the cylinder coincides with the regions of both the breakdown bubbles.
This implies that the vortex breakdown is characterized by the negative helicity in the vicinity of the axis.
Hence, it can be concluded that the topology of the vortex breakdown bubble is determined by the orientation of the velocity vector in the $rz$ plane and the vorticity vector in the same plane,
which in turn is characterized by $h_{r,z}$.
This decomposition is valid as far as the flow remains axisymmetric even in the unsteady regime.
The Figs. \ref{3000_t1} and \ref{3000_t3} show the comparison of the breakdown bubble against the negative contours of $h_{r,z}$ in $rz$ plane at two different instants for $\Gamma=2.5$ and $Re=3000$.
It can be seen that at each instant, the topology of the breakdown bubble is determined by $h_{r,z}$.
\begin{figure}
\centering
  \begin{subfigure}[b]{0.33\columnwidth}
  \centering
      \centering
      \includegraphics[width=0.5\linewidth]{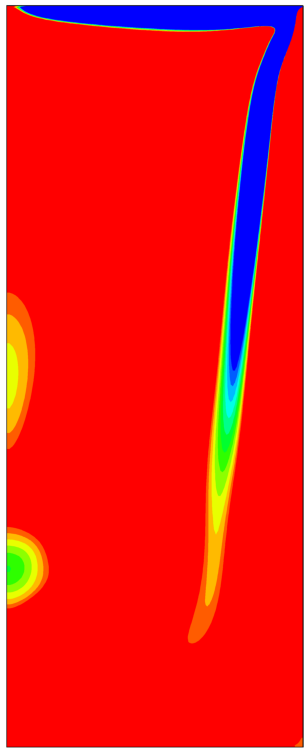}
      \caption{}
      \label{hrz_new}
    \end{subfigure}\hfill
    \begin{subfigure}[b]{0.33\columnwidth}
      \centering
      \includegraphics[width=0.5\linewidth]{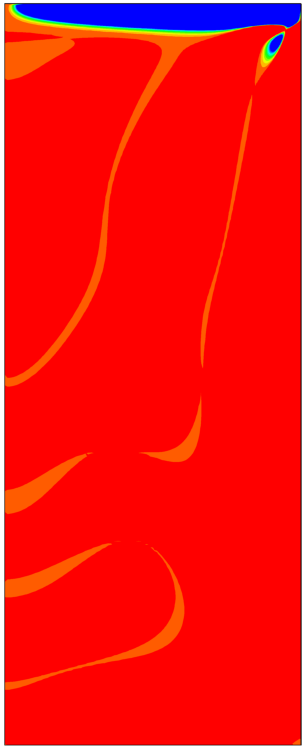}
      \caption{}
      \label{hrz1}
    \end{subfigure}\hfill
    \begin{subfigure}[b]{0.33\columnwidth}
      \centering
      \includegraphics[width=0.5\linewidth]{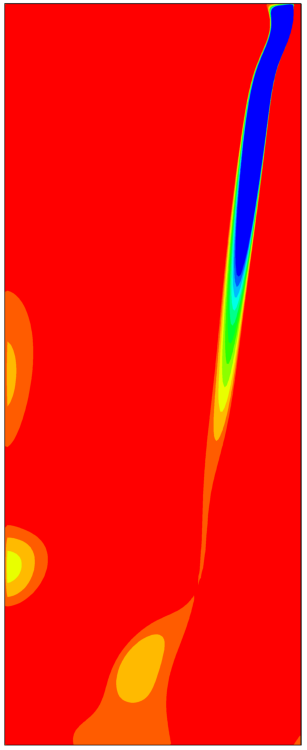}
      \caption{}
      \label{hrz3}
    \end{subfigure}\hfill
    \begin{subfigure}[b]{0.56\columnwidth}
      \centering
      \includegraphics[width=0.7\linewidth]{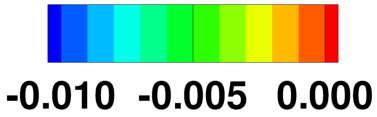}
      \label{}
    \end{subfigure}
\caption{Contours of negative values of (a) $h_{r,z}$, (b) $h_{r}$ (term $I$ of equation \ref{hrz_eqn}), and (c) $h_{z}$ (term $II$ of equation \ref{hrz_eqn}) in $rz$ plane for $\Gamma=2.5$ and $Re=2200$.}
\label{hrz_contri}
\end{figure}
We look at the contributions from the individual terms in the equation \ref{hrz_eqn} which are plotted in the Fig. \ref{hrz_contri}.
The figure shows that the main contribution to the breakdown bubble comes from term $II$ of the equation \ref{hrz_eqn} as the term $I$ is negligible compared to this term.
This term $II$ is essentially the axial part, $h_{z}$, of $h_{r,z}$.
This implies that $u_{z}$ and $\omega_{z}$ alone determine the topology of the vortex breakdown bubble.

\paragraph*{Conclusion---}
Helicity density for vortex breakdowns has been analyzed numerically.
Negative azimuthal vorticity has been found to be generated even in the cases when there is no vortex breakdown.
We find that the helicity density is a good parameter to characterize the vortex breakdown.
Helicity is negative in the regime where the vortex breakdown bubble occurs, indicating that the breakdown bubble is a
different topology than the surrounding flow.
This further implies that while a thunderstorm is characterized by high helicity, its eye should be characterized by negative helicity.
In the axisymmetric limit, we have shown that the flow can be assumed as essentially two-dimensional with the azimuthal component
of the velocity advected as a passive scalar in this flow field.
In such case, we find that the topology of the breakdown bubble is entirely determined by the helicity in the $rz$ plane and $h_{z}$
contributes most towards the topology of the breakdown bubble.
This indicates that the structure of the flow is dependent on the mutual orientation of the two-dimensional velocity vector
and the in-plane vorticity vector, at least for the axisymmetric flows.
This study highlights the importance of the helicity in the vortex breakdown in a way which has not been discussed previously.
Even though, such a decomposition is possible when the flow becomes non-axisymmetric, the azimuthal component of the flow no longer remains passive
and the effect of the out of the plane component, $h_{\theta}$, of the helicity needs to be investigated separately.

\end{document}